# Running and tumbling with *E. coli* in polymeric solutions

A. E. Patteson[1], A. Gopinath[1,2], M. Goulian[3], and P. E. Arratia[1]

[1] *Department of Mechanical Engineering & Applied Mechanics,
University of Pennsylvania, Philadelphia, PA 19104*
[2] *Department of Physics & Astronomy, Haverford College, Haverford,
PA 19041*
[3] *Department of Biology, University of Pennsylvania, Philadelphia, PA 19104*

Run-and-tumble motility is widely used by swimming microorganisms including numerous prokaryotic eukaryotic organisms. Here, we experimentally investigate the run-and-tumble dynamics of the bacterium *E. coli* in polymeric solutions. We find that even small amounts of polymer in solution can drastically change *E. coli* dynamics: cells tumble less and their velocity increases, leading to an enhancement in cell translational diffusion and a sudden decline in rotational diffusion. We show that suppression of tumbling is due to fluid viscosity while the enhancement in swimming speed is mainly due to fluid elasticity. Visualization of single fluorescently labeled DNA polymers reveals that the flow generated by individual *E. coli* is sufficiently strong to stretch polymer molecules and induce elastic stresses in the fluid, which in turn can act on the cell in such a way to enhance its transport. Our results show that the transport and spread of chemotactic cells can be independently modified and controlled by the fluid material properties.

Flagellar propulsion of microorganisms is perhaps one of the earliest forms of motility [4, 5]. This flagellar propulsion plays an important role in various biological and ecological settings, such as the spread and control of diseases [6-9], transport in lakes and oceans [10] and the biodegradation of environmental pollutants [11]. Therefore, it is essential to understand the role of the ambient environment in mediating and influencing the motility of microorganisms. Many of these environments are liquid-like and contain particles, polymers or other macromolecules, which introduce non-Newtonian features to the fluid such as shear-thinning viscosity and elasticity. These so-called *complex* fluids can strongly affect the motility of microorganisms [12-15]. For instance, glycoproteins in the stomach mucus form a viscoelastic gel that offer an effective barrier against most parasitic microorganisms. Yet, the bacterium *H. pylori* excretes enzymes that transform the impenetrable gel into a viscous polymer solution, which enables swimming and ultimately leads to persistent infections [16]. In this case, the subtle interplay of cell activity and complex material properties has significant impact: *H. pylori* alone infects 50% of the world's population [8], and more generally, bacteria comprise 65% of human microbial infections [17]**.**

Many additional biological functions rely on the motion of living particles in complex fluids, including fertilization through sperm cells swimming within cervical mucus [18] and the transport of mucus in the human lungs by rhythmically beating cilia [19]. An emerging number of investigations reveal intricate (and sometimes contradictory) ways in which the fluid material properties affect the motility of microorganisms. For example, fluid elasticity has been found to either enhance [20-24] or hinder [13, 25, 26] microorganism's swimming speed depending on the details of the swimming kinematics and the generated flow fields. Recently, the effects of shear-thinning viscosity, a common attribute of many polymeric fluids, have been found to have little to no effect on swimming speed in experiments [27, 28] and theoretical studies [29]. In contrast, experiments with the bacterium *E. coli* indicate that the shear-thinning viscosity of semi-dilute polymer solutions can lead to an enhancement in swimming speed [28]. Together, these works highlight the subtle interplay between fluid material properties and swimming kinematics, which results in a striking and often unanticipated variety of outcomes.

In this manuscript, we focus on run-and-tumble motility, a general mechanism employed by many prokaryotic flagellated bacteria (e.g. *E. coli, S. marcescens,* and *V. alginolyticus*) and even some eukaryotic organisms such as the green algae *C. reinhardtii* [30]. This mechanism can be described as a repeating sequence of two actions: (i) a period of nearly constant-velocity straight-line translation (run) followed by (ii) a seemingly erratic rotation (tumble). This run and tumble series -- a hallmark of many swimming bacteria – ultimately dictates their spread and transport. Here, the transport is effectively described by a persistent random walk with an active effective diffusion coefficient. While the run and tumble mechanism has been widely studied in simple, water-like (i.e. Newtonian) fluids [31-34], many bacteria that employ this mechanism live in biological fluids that contain macromolecules and are not Newtonian. Since motility is directly linked to virulence [6, 9], understanding the role of fluid rheology on run-and-tumble dynamics and the overall spread of bacteria is therefore of much practical interest.

Here, the run-and-tumble motility of the bacterium *E. coli* is experimentally investigated in polymeric solutions using cell tracking methods and single molecule experiments. The bacterium *E. coli* is an archetypical model for studies of run-and-tumble dynamics [31, 34]. *E. coli* is known to thrive in the human digestive tract (a viscoelastic medium) and is a common agent for food poisoning [35]. We find that the presence of even small amounts of polymers in solution dramatically alters the cell motility: tumbling is suppressed and cells swim faster. By varying (i) the type of polymer, (ii) polymer molecular weight (MW) and (iii) polymer concentration, we show that fluid viscosity suppresses tumbles while fluid elasticity enhances swimming speed. We also show in single molecule experiments using fluorescently labeled DNA polymers that the flow field generated by *E. coli* is able to stretch initially coiled polymer molecules, and thus induce elastic stresses in the fluid. These changes in motility behavior, driven by the material properties of the ambient fluid, can have profound influences on transport and foraging of nutrients. Our results also suggest that tuning the material properties of the fluidic environment can control the spreading of bacteria.

We experiment with different types of polymeric fluids and a water-like buffer solution. Three main types of polymer molecules are used: poly-ethylene glycol (PEG, Sigma-Aldrich, MW = 8.0 x $10^{4}$, $R_g \approx 6$ nm), carboxy-methyl cellulose (CMC – a linear, flexible polymer, Sigma-Aldrich, MW = 7.0 x $10^5$, $R_g \approx 28$ nm) and xanthan gum (XG, Sigma Aldrich, MW 2.0 x $10^6$, $R_g \approx 600$ nm), where $R_g$ is the polymer radius of gyration. We note that radius of gyration of the polymer molecules range from 6 nm to 600 nm. This range is comparable to the width of a single *E. coli* flagellum (approximately 20 nm) but smaller than total effective length of the bacterium (body plus flagellar bundle) of approximately 7 $\mu$m [31]. We varied CMC concentrations from 10 to 500 ppm – significantly below the overlap concentration of $10^4$ ppm – to diminish the role of polymer-polymer interactions and avoid the presence of polymer networks. To discriminate between the roles of elasticity and shear-thinning fluid properties, we also use CMC of different molecular weights (9.0 x $10^4$, 2.5 x $10^5$, and 7.0 x $10^5$) as well as solutions of xanthan gum, a semi-rigid polymer. Xanthan gum solutions exhibit shear-thinning viscosity and elasticity. By adjusting the polymer concentration and MW, we make fluids of desirable viscosity ($1 \leq \mu < 20$ mPa s) and elasticity (fluid relaxation time $\lambda$ up to 50 ms [36]). Finally, Newtonian fluids are prepared using (i) a water-like buffer solution of 67 mM NaCl in water and (ii) PEG aqueous solutions. The concentration of PEG in solution varies from 1.3 to 3.5 % by weight. All PEG solutions display Newtonian viscosity. See SI1 for rheology details.

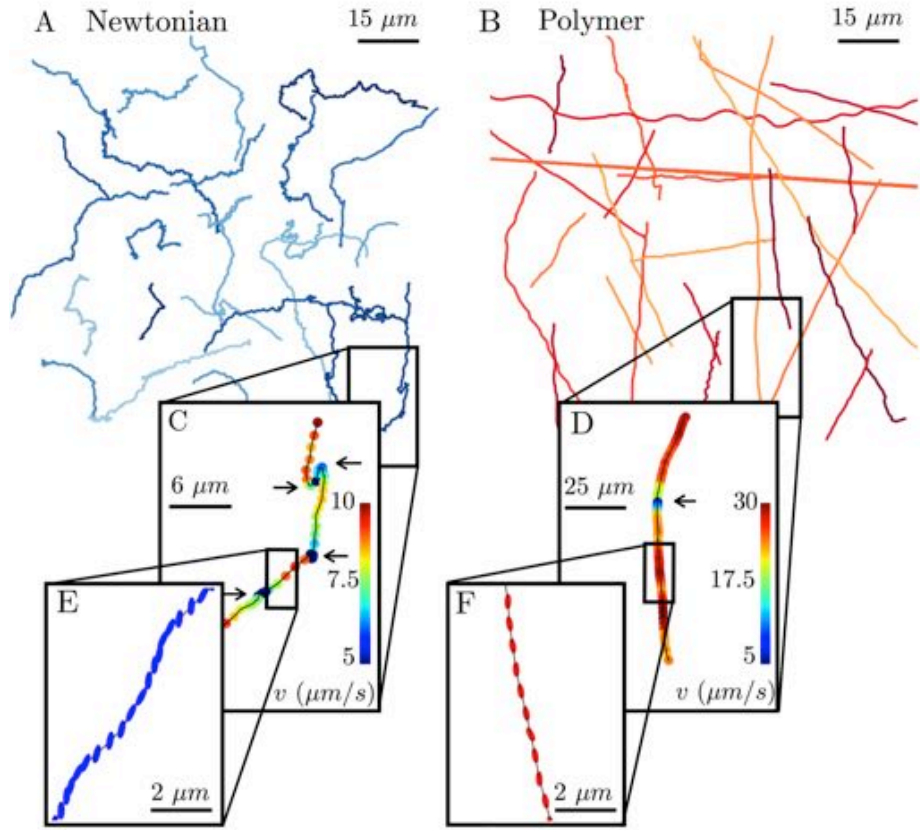

**Figure 1**: Kinematics of swimming *E. coli* cells in both Newtonian and viscoelastic fluids. (A) Trajectories of *E. coli* cells in buffer ($\mu$ = 1.0 mPa · s) and (B) in polymeric solution (CMC MW = 7 x $10^5$, *c* = 500 ppm, *c\** =$10^4$, $\mu$ = 19 mPa · s). Cells in polymer solution move remarkably straighter compared to cells in buffer. Sample cell trajectories in (C) buffer and (D) polymeric solutions exhibit run-and-tumble i. e. nearly straight lines connected at random angles (tumbles denoted by arrows). (E) The cell body orientation $\phi$ oscillates or 'wobbles' in the buffer solution. (F) Wobbles diminish in the polymer solution.

Our experimental protocol consists of directly observing *E. coli* cells suspended in thin fluid films (Methods). We track the orientation of representative cell bodies via the angle $\phi$, defined as the angle made by the unit vector aligned with the major axis of the elliptical cell body **p** and the *x* axis, $\cos \phi = \boldsymbol{p} \cdot \boldsymbol{e_x}$. The orientation of the trajectory is tracked using the angle $\theta$ defined by $\cos \theta = (\boldsymbol{r} \cdot \boldsymbol{e_x})/|\mathbf{r}|$, where $\mathbf{e}_x$ is the unit vector aligned with the *x*-axis.

Representative *E. coli* trajectories in buffer (Newtonian) and carboxy-methyl cellulose (CMC, MW = 7.0 x $10^5$, *c* = 500 ppm) solutions are shown in Fig.

1(a) and 1(b), respectively. In buffer solution, cells swim in various directions, executing a random walk and frequently change direction, typical of the run-and-tumble mechanism [31]. Figure 1(b) reveals a very different behavior. If we replace the Newtonian fluid by the CMC solution, the cell paths are smoother and straighter, exhibiting changes in direction less frequently. We further illustrate these changes in swimming behavior by examining sample trajectories (time interval of 2 seconds) in buffer (Fig. 1c) and CMC (Fig. 1d) solutions. We identify tumbles (arrows in Fig. 1c and d) in the sample trajectories by tracking sudden changes in direction and simultaneous drops in speed. Surprisingly, we find that cell trajectories in the CMC solutions are nearly devoid of tumbles compared to the buffer case. Figure 1(e,f) shows the instantaneous cell body orientation $\phi$ during sample trajectories for a fixed distance (~10 µm). The data shows that $\phi$ is also strikingly different for cells swimming in polymeric solutions. Figure 1(e) shows that, in buffer solution, the orientation of the cell body oscillates significantly along its path. These two-dimensional lateral oscillations of the cell body, known as "wobbling", are projections of the cell's three-dimensional helical trajectory [37, 38]. In the CMC solution, however, this oscillation ( wobbling) significantly diminishes (SI Movie 1), and $\phi$ remains relatively constant (Fig. 1f). This hints to a change in the *E. coli* swimming kinematics such as the pitch or angle of the cell helical path.  Overall, the results shown in Fig. 1 indicate that the presence of even small amounts of polymer in liquids can significantly affect the motility of microorganisms and, in the case of *E. coli*, suppresses tumbles and body oscillations.

To quantify the above observations, we calculate the *E. coli* instantaneous velocity $v$ and its magnitude $|v|$ as a function of time from the tracking data. The velocity vector is defined over a time interval of $\Delta t$ =1/15 s, which is large enough to average over $\phi$ yet small enough to define an average swimming orientation $\theta$ between tumbles (a). Figure 2b shows examples of velocity magnitudes $|v|$ as a function of time for buffer and CMC solutions ($c$ = 500 ppm). The data shows that cells swimming in CMC solutions execute tumbles (denoted by arrows) less frequently than in buffer (Newtonian) solutions. Here, the cell in buffer tumbles 5 times in the span of 6 seconds. In contrast, the cell in polymeric solution (CMC) tumbles only twice in the same time span.

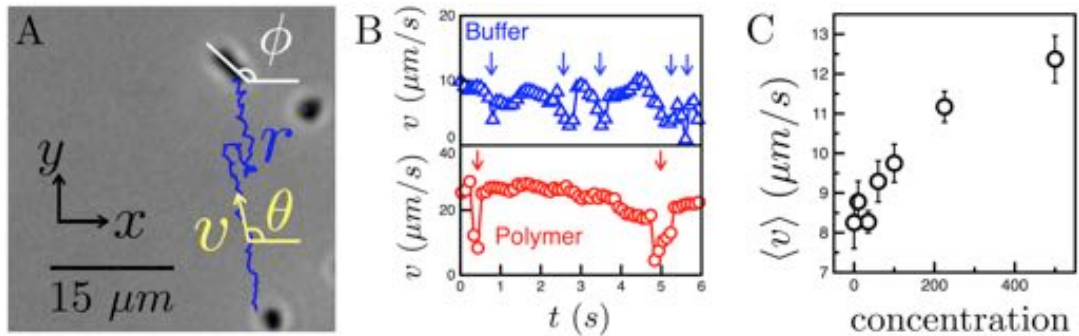

**Figure 2**: Swimming speeds of *E. coli* in buffer and polymeric solutions. (A) Velocity $v$, average cell orientation $\theta$, and instantaneous cell body orientation $\phi$ are defined as shown. (B) Temporal variations in the cell body speeds in buffer and polymeric solutions ($c$ = 500 ppm) reveal tumbles (arrows) via sudden drops in $v$. The cell in buffer swims at a lower velocity and tumbles more frequently compared to the cell in the polymer solution. (C) The mean cell velocity increases from 8.3 to 12.4 µm/s with increasing polymer concentration.

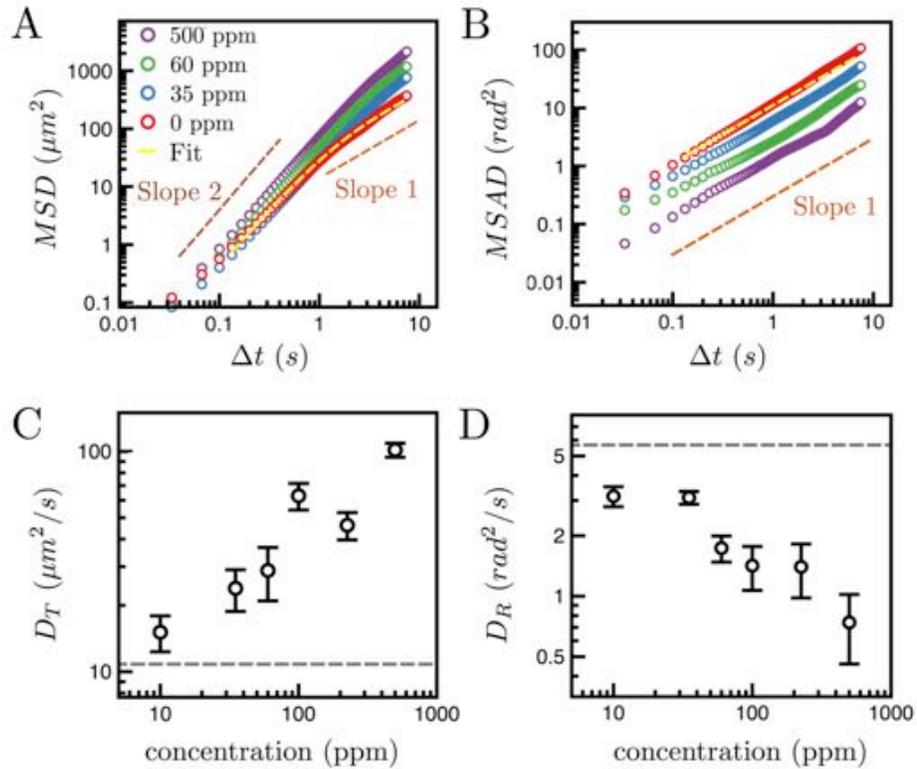

**Figure 3:** Statistical measures characterizing cell trajectories. (A) The mean-square displacement for cells in buffer and CMC solutions (concentration $c$ = 0, 35, 60, 100 ppm, MW = 7 x $10^5$). At short times, $\Delta t \ll \tau_R$, where $\tau_R$ is the mean run time, the cell motion is ballistic, and $MSD \propto (\Delta t)^2$. At longer times, $\Delta t \gg \tau_R$, the cell motion is diffusive and $MSD \propto \Delta t$. As $c$ increases, the magnitudes of the MSD curves increase. (B) The mean-square angular displacement of cells in buffer and polymeric solutions increases linearly over time, indicating diffusive reorientations. (C) The translational diffusion coefficient increases from 10.8 to 101.6 µm²/s as $c$ increases. The result for buffer ($c$ = 0 ppm) provides a reference (dashed line). (D) The rotational diffusion coefficient $D_R$ decreases from 5.6 to 0.7 rad²/s as $c$ increases, reflecting suppressed tumbling in polymeric solutions.

The sample velocity records in Fig. 2(b) show that the *E.coli* swims faster in CMC solution (25 µm/s) than in the buffer (10 µm/s) even though the CMC solution has a viscosity that is over an order of magnitude ($\mu \approx 20$ mPa $\cdot$ s) larger than that of the buffer ($\mu \approx 1$ mPa $\cdot$s). In fact, Fig 2(c) shows that the mean instantaneous cell velocity $\langle v \rangle$ (averaged over hundreds of individual cells) increases with polymer concentration from about 8.3 µm/s in buffer solution to 12.4 µm/s in CMC solutions ($c = 500$ ppm); the speed in buffer is consistent with previous measurements [39]. This enhancement in $\langle v \rangle$ with polymer concentration is somewhat counterintuitive, since the viscosity increases as polymer is added to the fluid (SI1). We note that in a Newtonian fluid the viscous torque on the cell flagella bundle $\tau_b$ is proportional to $\mu\omega$, where $\omega$ is the bundle rotation rate. For *E. coli* swimming at constant motor torque $\tau_m$ [34], the torque balance yields $\tau_m \sim \tau_b$, and thus $\tau_m$ is also proportional to $\mu\omega$. In highly viscous environments corresponding to swimming at low Reynolds number, Stokes equations hold and thus the speed varies with the frequency $v \propto \omega$ [14, 37, 40, 41]. Therefore, as viscosity increases, the bundle rotation rate $\omega$ and correspondingly the forward velocity should decrease as $\mu^{-1}$. The increase in average velocity $\langle v \rangle$ with polymer concentration is thus unexpected.

Similar increases in cell velocity with polymer concentration have been previously reported [40, 42]. It has been argued that cell velocity is augmented by the presence of a gel-like network which exerts an anisotropic viscous drag on the cell [43]. In our experiments, however, the CMC (polymeric) solutions are considered dilute ($c \lesssim 5\%$ of the overlap concentration) in the sense that polymer networks are not present. Thus the anisotropic viscosity argument given by [43] does not explain our results. More recently, Martinez *et al*. [28] argued that shear-thinning viscosity of semi-dilute polymeric solutions was responsible for enhancing the *E. coli* swimming velocity. Here, we will show an alternative explanation. Namely, the increase in swimming speed can also be due to extra elastic stresses.

Next, we quantify the effective translational ($D_T$) and rotational ($D_R$) diffusions of swimming *E. coli* by computing the mean-squared displacement (MSD) and the mean-squared angular displacement (MSAD) from tracking data, as shown in Fig. 3(a) and 3(b). The mean-squared displacement is defined as $MSD(\Delta t) = \langle |\boldsymbol{r}(t_0 + \Delta t) - \boldsymbol{r(t_0)}|^2 \rangle$. For a random walk, the MSD is $4D_T\Delta t$ in two dimensions, where $D_T$ is the effective translational diffusion coefficient. For a swimming *E. coli* at short time intervals, the MSD is proportional to $\Delta t^2$ (Fig. 3a), indicating the cells swim ballistically during a run. For times much larger than the mean run time $\tau_R$, the cells tumble, decorrelating their motion. Thus for very large $\Delta t \gg \tau_R$, the motion is diffusive as seen in Fig 3 (a).

For *E. coli*, the dynamics can be captured using the relationship $MSD(\Delta t) = 4D_t\Delta t(1 - e^{-\Delta t/\tau})$, where $\tau$ is a typical crossover time marking the transition from ballistic to diffusive motion (see SI2 for details). The crossover time depends on the mean run time $\tau_R$ corrected by a factor that accounts for the mean cosine of the turning angle $\alpha$ such that $\tau = \tau_R/(1-\alpha)$ [44]. The MSD is proportional to $4D_T(\Delta t)^2/\tau$ for $\Delta t \ll \tau_R$ and to $4D_T\Delta t$ for $\Delta t \gg \tau_R$. By fitting this relationship to the MSD data in Fig. 3(a), we find that the translational diffusion coefficient $D_T$ increases significantly from 10.8 to 101.6 µm²/s as polymer concentration (and viscosity) increases (Fig. 3c). The crossover time $\tau$ also increases with polymer concentration from 0.9 to 4.8 s (SI3). This suggests an enhancement in mean cell run time, consistent with the observed suppressed tumbling in polymer solutions (Fig. 1).

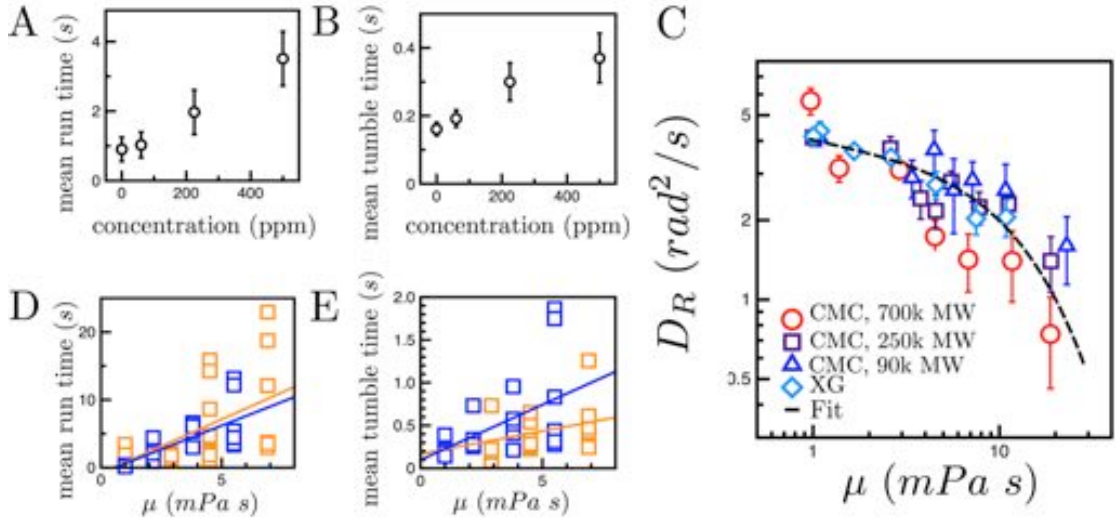

**Figure 4:** Viscosity suppresses tumbling. (A) The mean run time increases from 0.95 to 3.51 s as the CMC polymer concentration *c* increases (MW = 7 x 10⁵). (B) The mean tumble time also increases with *c* from 0.16 to 0.37 s. (C) The rotational diffusion coefficient $D_R$ decreases with viscosity for the CMC and XG solutions, indicating that suppressed tumbling is nearly independent of MW or molecule, and is captured by proposed model. (D) The mean run and (E) mean tumble times for individual tethered cells in Newtonian (PEG, blue squares) and viscoelastic (CMC, orange squares) fluids as a function of $\mu$ (Movie 2). Lines correspond to regression analysis (Methods).

Next, the *E. coli* rotational diffusivity is investigated by calculating the mean-squared angular displacement, defined here as $MSAD(\Delta t) = \langle |\theta(t_0 + \Delta t) - \theta(t_0)|^2 \rangle$. We use the cell orientation $\theta$ to construct the MSAD data, which is shown in Fig. 3b. Then, the data is fitted to $MSAD = 2D_R\Delta t$ in order to obtain the effective rotational diffusion coefficient $D_R$. For the buffer solution case, $D_R$ is approximately 5.6 rad²/s (Fig. 3d). For cells swimming in CMC solutions, the values of $D_R$ diminish to 0.7 rad²/s (c=500 ppm). The decrease in rotational diffusivity is also consistent with the appearance of nearly straight trajectories in polymeric solutions (Fig. 1b).

To connect the time-averaged statistical quantities of swimming *E. coli* to their instantaneous kinematics, we measure the mean run and tumble times as shown in Fig. 4(a) and 4(b). Mean run time is defined as the time intervals between successive tumbles, identified here by rapid drops in velocity (Fig. 2b). We find as polymer (CMC) is added to the fluids, the run times increase from approximately 0.9 to 3.5 s (Fig. 4a). This enhancement in run time is consistent with the nearly straight trajectories (c.f. Fig. 1b) and the reduction in rotational diffusivity in polymeric solutions. The mean tumble times (Fig. 4b) are defined as the mean time intervals between runs. This quantity also increases (from 0.2 to 0.4 s) with polymer concentration. This observed increase in both run and tumble times is in marked contrast to chemotactic cells in chemical gradients in which run times increase but tumble times remain constant [31]. Thus, the *E. coli* biochemical signaling network cannot solely explain our results, suggesting that the fluid rheology is affecting the cell motility behavior. We note that the mean run and tumble times are consistent with previous measurements [32].

In order to investigate which fluid properties contribute to the changes in *E. coli* run and tumble times, we measure the rotational diffusivity $D_R$ in fluids with varying rheological properties. We note that $D_R$ for an *E. coli* is inversely proportional to the mean time $\tau_R$ (see SI4 for details) [44]. These fluids are polymeric solutions of CMC of different molecular weight (MW) and XG. Figure 4c shows the cell rotational diffusivity $D_R$ as a function of fluid viscosity $\mu$. The data clearly shows that, for all solutions, $D_R$ decreases with $\mu$. The agreement in the data for multiple fluids and two types of polymers indicates that $D_R$ is independent of the variations in elasticity and shear-thinning properties (SI1). The decrease in $D_R$, which scales as $D_R \sim 1/\tau_R$, thus indicates an increase in run times $\tau_R$, and the collapse in Fig. 4(c) strongly suggests that $\tau_R$ predominately depends on fluid viscosity.

To better understand the observed enhancement in run and tumble times with $\mu$, we perform experiments in which the run and tumble states of the cell can be directly visualized by the rotation of tethered *E. coli*. Sticky-flagellated mutant *E. coli* can tether to glass slides [34]. The resulting counter-clockwise (CCW) or clockwise (CW), rotation of cell bodies corresponds to the run or tumble state of the motor, respectively (SI Movie 2, Methods). Figures 4(d) and 4(e) show the mean run and tumble times as a function of viscosity for individual cells in viscoelastic (CMC) and Newtonian (PEG) fluids. The mean run and tumble times tend to increase with viscosity for both fluids. Linear regression analysis reveals that this increase is statistically equivalent in the Newtonian and viscoelastic fluids (Methods). The tethering results bolster our observations that the changes in *E. coli* run and tumble times are mainly due to changes in viscous stresses. We propose that as viscous stresses increase, the mechanical (viscous) load on the cell also increases which in turn affects the cell motor switching rates between run and tumble states. Previous experiments have in fact shown that mechanical loading can significantly affect motor switching rates [45, 46], where mechanical loads were introduced by attaching latex beads to the flagellar stubs.

To interpret these results (Fig. 4), we suggest a minimal model valid at high loads (as in our experiments) that treats motor switching as an activated process with rates controlled by effective energy barriers that need to be overcome for potential tumbles to occur [46, 47]. In the absence of external loading, the motor switching rate $k^*$ depends on the chemical binding rate of a signaling molecule Che-Y to the cell motor. Assuming that viscous drag on the cell flagella presents an additional energy barrier to switch from one state to the other, the switching rate $k$ is modified to $k \propto k^* \exp\left(-\beta M/k_B T\right)$, where $M$ is a characteristic external torque generated by viscous drag on the flagella and $\beta$ is a characteristic angle determined by the internal details of the coupling between the flagella and motor necessary to switch states (Methods). As fluid viscosity increases, the torque $M$ increases, and the switching rate decreases by a factor $\exp\left(-\beta M/k_B T\right)$, consistent with the observed enhancement in run and tumble times (Fig. 4a-b,d-e).

As the motor switching rates diminish with increased viscous loading, the cell rotational diffusion $D_R$ is suppressed (Fig. 4c). This decrease in $D_R$ may be interpreted as follows. The rotational diffusivity of an *E. coli* is a sum of its Brownian rotational diffusivity $D_R^0$, arising due to passive thermal motion, and its active rotational diffusivity due to tumbles [44]. The Brownian rotational diffusion of a particle is $D_R^0 = k_B T/f_0 \mu$, where $f_0$ is the geometry-dependent resistivity according to the Stokes-Einstein relationship. Assuming that the *E. coli* body is an ellipsoid (2 µm long and 1 µm wide), $f_0$ is approximately $9.45\,\mu m^3$ [48]. If the mean run time increases as $\exp\left(\beta M/k_B T\right)$ and the torque $M$ is proportional to viscosity $\mu$, then the rotational diffusion coefficient follows $D_R = D_R^0 + A^* k^* e^{-\beta M/k_B T} = \frac{k_B T}{f_0 \mu} + A e^{-B\mu}$. By fixing $f_0$ to 9.45 µm³ and temperature $T$ to 22°C, we fit this equation to the data in Fig. 4c and obtain $A$ = 3.85 rad²/s and $B$ = 68.3 (Pa s)⁻¹. The parameter $A$ is a constant rotational diffusion based on the cells intrinsic motor switching rate $k^*$. The parameter $B$, defined here as $B = \beta M/k_B T \mu$, corresponds to a motor torque $M = 650$ pN nm in water [37] and a characteristic angle $\beta = 0.025°$ (Methods). The model seems to capture the main features of the $D_R$ versus viscosity data and further supports the idea that the decrease in rotational diffusion of swimming *E. coli* is due mainly to mechanical loading of the motor via viscous drag.

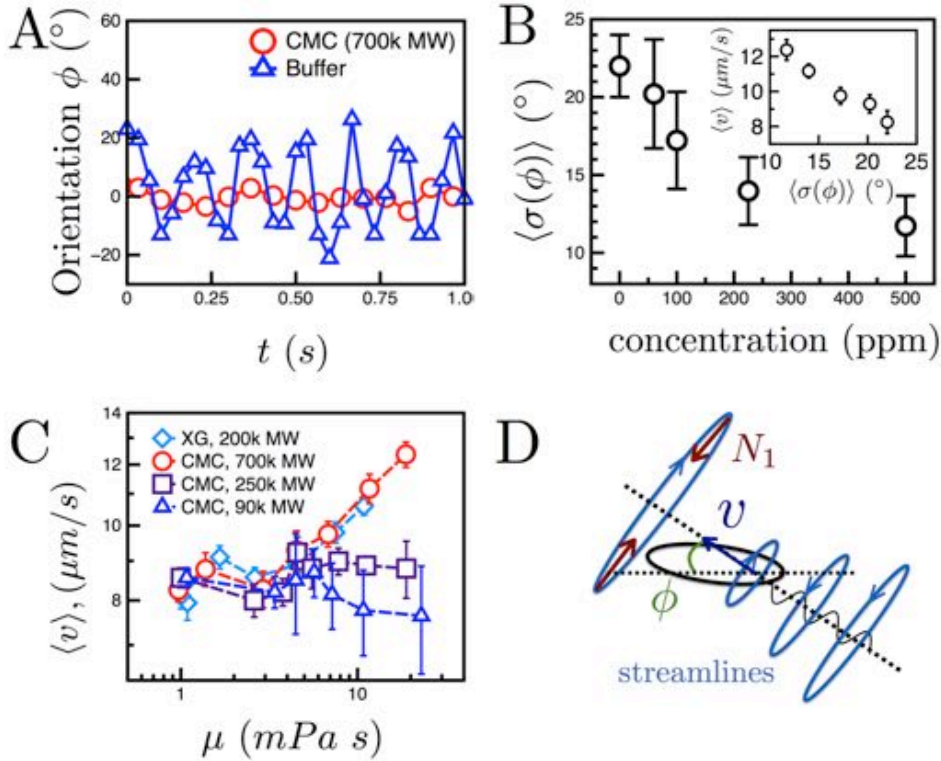

**Figure 5:** Elasticity suppresses wobbling while increasing cell velocity. (A) The body orientation $\phi$ versus time for a cell in buffer and polymer solutions (CMC, MW = 7.0 x $10^5$, $c$ = 500 ppm). In buffer, the cell wobbling amplitude is significantly larger than in the polymer solution (SI Movie 3). (B) The degree of wobbling, $\langle\sigma(\Phi)\rangle$, decreases from 22.0 to 11.7° as the CMC polymer concentration increases. (Inset) Mean cell velocity decreases with $\langle\sigma(\phi)\rangle$, illustrating that cells which wobble less swim faster. (C) Mean cell velocity $\langle v \rangle$ versus viscosity $\mu$ for solutions of CMC of varying molecular weight and XG. The velocity increases with $\mu$ for the largest MW of CMC but remains nearly constant in the lowest MW. (D) As *E. coli* swim, they generate a fluid flow with curved streamlines [1]. This shear can stretch polymers, producing first normal stress differences $N_1$. Under these curved streamlines, a volume force ($N_1/r$) points inward to the cell body, suppressing wobbling, and allowing cells to translate at higher $v$.

Next, we investigate the enhancement of cell velocity with increasing polymer concentration (Fig. 2c). The increase in polymer (CMC) concentration leads to an increase in fluid viscosity $\mu$ and elasticity (SI1). Here we argue that the observed increase in cell velocity is due to elastic stresses, which suppress cell wobbling (as shown in Fig. 1 (e,f)) and allow the cells to translate more efficiently. A decrease in *E. coli* wobbling has been previously observed in polymeric solutions [37], but the connection to cell swimming speed has not been made. We begin by tracking the orientation of the cell body $\phi$ relative to the direction of its trajectory in buffer and CMC ($c$ = 500 ppm, MW = 7.0 x $10^5$) solutions (Fig 5a). The estimated wobble angles are approximately 20° and 5° in buffer and CMC solutions, respectively. There is, therefore, a significant suppression of wobbling as polymer concentration is increased. We can further characterize this suppression by computing the mean standard deviation of $\phi$, $\langle \sigma(\phi) \rangle$, over many cells. This quantity $\langle \sigma(\phi) \rangle$ characterizes the degree of wobbling. Figure 5b shows that the quantity $\langle \sigma(\phi) \rangle$ decreases from 22.0° to 11.7° with increasing CMC polymer concentration. The decrease in $\langle \sigma(\phi) \rangle$ signifies a change in the cell swimming kinematic or stroke. Also, Fig. 5 (b, inset) shows that the cell velocity $\langle v \rangle$ is inversely proportional to the degree of wobbling $\langle \sigma(\phi) \rangle$; that is, a suppression in wobbling leads to an increase in cell velocity.

In order to distinguish between elastic and viscous effects, we measure *E. coli* mean cell velocity $\langle v \rangle$ and degree of wobbling $\langle \sigma(\phi) \rangle$ in CMC and XG solutions. Figure 5(c) shows $\langle v \rangle$ as a function of fluid viscosity $\mu$ for CMC solutions of varying MW and a XG solution. While $\langle v \rangle$ increases with $\mu$ for the highest molecular weight CMC and XG solutions, the relative enhancement in $\langle v \rangle$ diminishes as the CMC molecular weight (and thus elasticity) decreases. This is evident if one considers $\mu$ = 11 mPa · s, where $\langle v \rangle$ clearly decreases with the MW of CMC. This observation suggests that *E. coli* swimming speed $\langle v \rangle$ is not a function of fluid viscosity. Also, it appears that shear-thinning effects are negligible since the values of $\langle v \rangle$ for the highest molecular weight CMC (weakly shear-thinning, power law index = 0.7) and XG (strongly shear-thinning, power law index = 0.5) solution in Fig. 5(c) are indistinguishable. The increase in $\langle v \rangle$ with CMC molecular weight (MW) is also consistent with a simultaneous decrease in wobbling (SI5). We conclude that the suppression of cell wobbling due to fluid elasticity results in an increase in cell swimming velocity $\langle v \rangle$.

What may cause fluid elasticity to suppress wobbling and thereby increase? We suggest a mechanism supported by our experimental observations by which this is accomplished. As a single *E. coli* swims through a fluid, it generates a flow with curved streamlines [1] due to the rotating flagella and the concomitant counter-rotation of its body, as shown schematically in Fig. 5d. In flow, shear can stretch flexible polymer molecules [49] (such as CMC) and generate first normal stress differences $N_1$. The combination of shear and curved streamlines produce a (volume) force $N_1/r$, which points inward in the radial direction ($r$). We propose that this force, which for an *E. coli* cell points into the cell body (Fig. 5d) and perpendicular to the cell's swimming direction, causes the cell body to align with the projected direction of motion. The resultant decrease in wobbling amplitude would ultimately change the form (shape) of the swimming trajectory and increase the cell swimming velocity $\langle v \rangle$. Thus, we propose that $\langle v \rangle$ increases with polymer concentration (Fig. 2c) primarily because of the appearance of the force $N_1/r$, which is able to suppress wobbling – and cells that wobble less inherently swim faster. The combination of reduced wobbling (and thus higher $\langle v \rangle$) with enhanced run times results in straighter, longer trajectories in polymeric solutions (Fig. 1b).

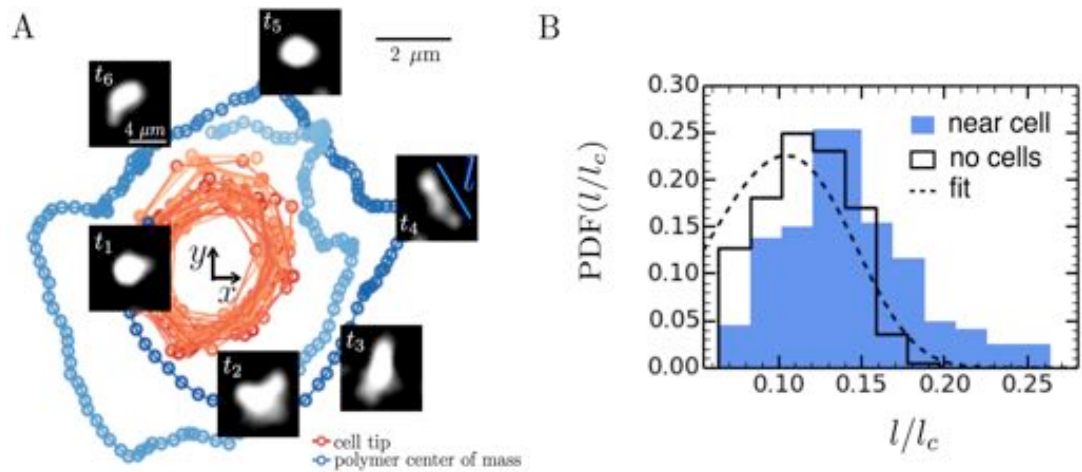

**Figure 6:** Polymer stretching by a tethered E. coli cell. (A) The tethered cell rotates counterclockwise (CCW) in a steady, circular trajectory. An untethered polymer molecule near the cell also rotates CCW due to hydrodynamic interactions with the cell (Movie 3). Sample configurations of the polymer ($\Delta t$ = 0.4 s) show extension and alignment with the flow. (B) The distribution of the normalized lengths $l/l_c$ for the polymer near the tethered cell (2.5 $\mu$m) is shifted to the right of the distribution in the absence of cells, suggesting that cell-generated flows stretch polymers and produce elastic stresses. The dashed line is the fit of the distribution for a self-avoiding chain at equilibrium (SI6) [2, 3].

This argument however is contingent on the expectation that swimming *E. coli* cells can actually generate flow fields strong enough to stretch polymer molecules and induce elastic stresses in a fluid. In order to gain further insight and verify that this is the case, we directly visualize the interaction of model polymer molecules and tethered *E. coli*. $\lambda$-DNA molecules are fluorescently stained and suspended in a buffer solution with mutant *E. coli* cells (Methods). These mutants contain sticky-flagella that can be tethered with ease and additionally also only `run'. As a result, there is a stable, three dimensional, time-dependent flow generated by the CCW-rotation of the tethered *E. coli* cell. We track the configurations of nearby DNA molecules over time, an example of which is shown in Fig. 6(a) (SI Movie 3). Also shown in Fig. 6(a) are the cell body and a nearby DNA molecule tracks over time. The sample snapshots ($\Delta t$ = 0.4 s) qualitatively show that the DNA molecular configuration evolves over time: it begins as a sphere, elongates and curves around the streamlines. These representative snapshots provide evidence that flows generated by moving *E. coli* are capable of stretching nearby polymer molecules, and thus induce elastic stresses in polymeric solutions.

In order to quantify the above observations, we measure the molecule (DNA) stretch length $l$ for two cases: (i) the absence of cells (i.e., no flow) and (ii) near a tethered cell, approximately 5 $\mu$m away from the cell. The distributions of DNA stretch lengths -- normalized by the $\lambda$-DNA contour length ($l_c$= 22.0 $\mu$m [50]) -- are shown in Fig. 6 (b) for both cases. In the absence of cells, the polymer molecules are in equilibrium and their configurations fluctuate randomly due to Brownian forces. The observed minimum $l/l_c$ in Fig. 6b corresponds to a length $l$ of approximately 1.4 $\mu$m, consistent with the length ($2R_g$) of a polymer with the inferred radius of gyration, $R_g \approx 0.7$ $\mu$m [50]. The peak in the distribution is followed by a rapid decay, which seems to follow the exponential decay of the theoretical end-to-end distance distribution (dashed line in Fig. 6b) of a self-avoiding polymer chain at equilibrium [2] and is also consistent with previous experimental measurements of $\lambda$-DNA [3, 50]. Compared to the DNA at equilibrium case (in the absence of cells), the length distribution of a polymer near a cell broadens and extends to higher values, reaching a maximum of approximately 7$R_g$ (Fig. 6b). For the DNA, this observed shift in the distribution corresponds to an applied force of approximately 4.5 fN (SI6) and is in reasonable agreement with expected viscous extensional forces generated by the tethered cell (Methods). The shift illustrates that the flow generated by the motion of the *E. coli* body in a fluid is indeed able to stretch polymer molecules beyond their equilibrium configuration.

To compare the DNA polymer extension by the tethered *E. coli* (Fig. 6) to the potential polymer extension by freely-swimming *E. coli* (Fig. 5), we estimate the Weissenberg number $Wi$ for both experiments. The Weissenberg number $Wi = \lambda \dot{\gamma}$, where $\lambda$ and $\dot{\gamma}$ are the fluid relaxation time and applied shear rates. We find that the Wi of the CMC and DNA polymer experiments are comparable, at approximately 13 and 8 respectively (SI7). This suggests that the CMC polymers near swimming cells exhibit similar stretching to the DNA polymer (Fig. 6) and may generate elastic stresses.

Our experiments highlight the complementary roles played by the elastic and viscous properties of complex fluids through which *E. coli* swim. For freely swimming *E. coli*, the stretching of nearby polymer molecules can lead to "extra" elastic stresses in the fluid [49], which act to align the cell body, reduce the degree of wobbling (Fig. 5b), and ultimately enhance cell velocity (Fig. 5). This increase in cell velocity with elasticity combined with the observed suppression of cell tumbles due to enhanced viscous loading (Fig. 4a) dramatically enhances the overall diffusivity and transport properties of bacterial cells (Fig. 3a,c) in fluids with small amounts of polymer.

Fluid properties such as viscosity and elasticity have been shown to significantly affect the motility of microorganisms. In this article, we investigated the effects of fluid material properties on the motility of *E. coli*. Using polymeric solutions of varying molecular weight, we found that the viscosity and elasticity can independently alter the swimming and transport of bacteria. In particular, we find that fluid viscosity suppresses cell tumbling, while fluid elasticity increases cell velocity. We also found that the flow generated by swimming bacteria influences the dynamics of polymers in solution, in such a way that the cells motility is enhanced. Direct visualization of individual tethered cells and nearby polymers reveals that cell-generated flows can indeed stretch and align polymer molecules, actively inducing local elastic stresses, which in turn act on the cell. These results complement recent simulations that predict unusual stretching in model polymers in the presence of multiple bacteria [51]. More broadly, our experiments highlight the need to consider the interactions between single polymer molecules and individual swimming microorganisms. These interactions and their emergent feedback mechanisms are crucial to many outstanding issues in engineering, biology, and medicine, such as the design of swimming micro-robots [15, 52] and the possible means to control biofilm formations [6, 7, 9, 17, 53]. Finally, our work emphasizes the need to study microorganisms in their *natural, non-ideal* environment, where complex material properties dramatically alter their macroscopic transport behavior.

## Methods

**Prepping and tracking cells suspended in thin film**
Suspensions of *E. coli* are prepared by growing the cells (wild type K12 MG1655) to saturation ($10^9$ cell/mL) in culture media (LB broth, Sigma-Aldrich). The saturated culture is gently cleaned by centrifugation and re-suspended in the fluid of choice at dilute concentrations ($5 \times 10^7$ cell/mL).

Experiments are performed in a thin fluid film by placing a 2-µl drop of cell-polymer/cell-buffer suspension in an adjustable wire frame and stretching the film to measured thickness 80 µm. The film interfaces are nearly stress-free which minimizes velocity gradients transverse to the film. *E. coli* are imaged with phase-contrast microscopy, and videos are taken at 30 frames per second. The positions of the cell body **r**(t) are gathered over time *t* via standard particle tracking techniques [54].

**Run and tumble times of tethered cells**

During the run or tumble states, the cell motor rotates in a counter clockwise (CCW) or clockwise (CW) direction, respectively, when viewed from behind. We use a sticky-flagellated mutant *E. coli* (strain MDG201) [34] to tether the cells to glass surfaces by their flagella. As the cell motor rotates, the body of the cell rotates about its tethered flagella in either a CCW or CW fashion, revealing the state of the motor. In SI Movie 2, sample tethered cells are shown in Newtonian fluids (solutions of PEG) and viscoelastic fluids (solutions of CMC). As the viscosity increases, the tethered cells exhibit two changes: (1) a decrease in rotational speed and (2) also an increase in both run (CCW) and tumble (CW) time intervals, measured from approximately 100 switching events.

For a cell in a Newtonian fluid, the torque on the motor is proportional to the frequency of rotation $\omega$ and the viscosity $\mu$. For cells that operate at constant torque [31, 34], an increase in viscosity should yield lower rotation rates, consistent with the observed decrease in rotation rates. In Fig 4E and 4E, we see that the mean run and tumble times of individual cells tend to increase with viscosity for both the CMC and PEG solutions. The increase in run and tumbles times are verified by linear regressions, which reveal positive correlations among the time intervals and viscosity. **Table 1** displays the slopes of the linear regressions. A t-test conducted at $\alpha$ = 0.05 ($t_c$ = 1.68) indicate that the slopes are statistically the same between the PEG and CMC solutions for the run time (t =0.9, p-value = 7 x $10^{-7}$) with viscosity and the tumbles times (t = 1.0, p-value =2 x$10^{-5}$) with viscosity. Furthermore, the presence of elasticity in the CMC does not significantly alter the run and tumble times. Instead, the increase in run and tumble times of tethered cells can statistically be accounted for by viscosity alone.

**Fluorescently-stained DNA molecules**

We fluorescently stain $\lambda$-DNA (MW = 3x$10^7$) polymers to visualize the interaction of tethered cells with individual polymer molecules. Suspensions of $\lambda$-DNA are prepared by heating $\lambda$-DNA stock solution at a temperature of 65 ℃ for 10 min and then quenching the sample in an ice bath for 3 minutes. The DNA molecules were stained with YOYO-1 iodide at a dye to base pair ratio of 1:4 and left to incubate at room temperature for one hour. The stained molecules were suspended in TE buffer with 4% (v/v) $\beta$-mercaptoethanol, which reduces the amount of photo-bleaching. The final concentration is 0.10 $c^*$, where $c^*$ = 40 µg/mL.

The fluorescently stained $\lambda$-DNA polymer molecules are suspended in a buffer solution with mutant *E. coli* cells: These mutants, strain PL4, contain the sticky-flagella for tethering and also always `run'. Once a tethered cell is identified using bright field microscopy, the polymer molecules around the cells are visualized with fluorescence microscopy (SI - Movie 3).

**Model for tumbling rates**

The *E. coli* motor is a rotary motor comprised of the flagellar hook and several rings of proteins and is driven by an ion gradient across the cell membrane [34]. The flow of protons through the motor induces conformational changes in the stator proteins, which generate a torque on the rotor. The binding of a protein molecule Che-Y to the cell motor induces a conformational change of the motor, thereby promoting the switching of the motor direction from CCW to CW and initiating a tumbling event. When Che-Y molecules unbind, the motor regains its original conformation and reverses direction again.

Duke *et al.* [47] proposed a thermal isomerization model to describe the switching dynamics in the absence of an external load. In this model, the motor switching rate is proportional to $e^{-\Delta G/kT}$, where $\Delta G$ is the energy difference between the free energy of the barrier and the energy of the CCW or CW state. The binding of Che-Y molecules to the motor lowers the free energy barrier, setting the internal switching rate $k^*$ [34, 35, 46, 47]. We propose that in a viscous fluid, the motor experiences a mechanical load due to viscous drag on the flagella. In order to reverse the motor rotation direction, the motor must overcome this viscous torque $M$. We hypothesize that this effect results in an additional energy barrier that has to be overcome for an attempted switching event to be ultimately successful. The height of this barrier may be estimated as the product of an external fluid resistive torque (and therefore external viscosity) and an internal state variable related to motor configurations, a characteristic angle $\beta$. With these simplifications, the net motor switching rate becomes $k^* e^{-M\beta/kT}$ and thus decreases with viscous torque on the flagella. Using this model, we predict the rotational diffusion of *E. coli* cells as a function of viscosity as $D_R = D_R^0 + A^* k^* e^{-\beta M/k_B T} = \frac{k_B T}{f_0 \mu} + A e^{-B\mu}$, where the mechanical loading is due to viscous stresses on the motor. The parameter $B$, defined as $B = \beta M/k_B T \mu$, corresponds to a motor torque $M = 650$ pN nm in water [37] and a characteristic angle $\beta = 0.025°$. This characteristic angle reflects the orientational change in the configuration of a stator protein subunit during a switching event [46]. Since the flagellar motor contains many stator protein subunits, one should more generally interpret $\beta$ as a weighted angle and $\beta M$ as a weighted average amount of work performed by the stators to switch the motor.

**Estimate of force generated from tethered cells**

The tangential flow field around a sphere of radius $a$ rotating about an axis with angular frequency $\omega$ and evaluated in its mid-plane is given by $v_\theta(r) \sim a\omega \, (^a/_r)^2$. Assuming the DNA molecule is at distance $R$, the velocity gradient in the radial direction can be estimated and the shear rate is roughly given by $|\dot{\gamma}| \sim 2(a/R)^3 \omega$. The actual shear rate differs from the order of magnitude estimate due to the shape of the bacterial cell and the presence of the wall. Using $a = 0.7 \, \mu$m, $R = 2.4 \, \mu m$ and $\omega = 0.9(2\pi)s^{-1}$, we find $|\dot{\gamma}| \sim 2.8 \, s^{-1}$. The polymer stretches from the small deviation from the streamline on which the center of mass moves. Balancing the lateral (cross streamline) extension of the blob with radius of gyration $R_g$ and persistence length $L_p$ of approximately 50 nm with Brownian forces, we estimate the flow induced force due to the fluid of viscosity $\eta$ extending the polymer as $F = \zeta\dot{\gamma} \, (^{2Tk_B}/_{k_\perp})^{1/2}$ where $\zeta = 6\pi\eta R_g$ and $k_\perp = {}^{Tk_B}/_{L_p l_c}$ are the viscous drag coefficient and effective polymer stiffness respectively. Plugging in values we find $F \sim 49$ fN –this estimate is an upper limit.

**Acknowledgements**

We thank N. Keim, D. A. Gagnon, S. Sakar, P. Olmsted, T. Lubensky and P. Purohit for insightful discussions as well as D. Wong, J. Guasto and G. Juarez for experimental assistance. We also acknowledge funding from NSF-CBET-1437482 and NSF DMR-1104705.

**Author contributions**

A. E. Patteson performed experiments, analysed data, and wrote manuscript. A. Gopinath analysed data, developed models, and edited manuscript. M. Goulian guided experiments with *E. coli* and mutants, and advised data analysis. P. E. Arratia designed experiments, advised data analysis and models, and edited manuscript.

**Additional information**

Supplementary information is available in the online version of the paper. Correspondence and requests for materials should be addressed to P. E. A.

**Competing financial interests**

The authors declare no competing financial interests.

**Tables**

|  | slope | $t_c$ | t | p-value |
|---|---|---|---|---|
| CMC run time | 1.59 | 1.68 | 0.86 | $7 \times 10^{-7}$ |
| PEG run time | 1.11 | 1.68 | 0.86 | $7 \times 10^{-7}$ |
| CMC tumble time | 0.053 | 1.68 | 1.04 | $2 \times 10^{-5}$ |
| PEG tumble time | 0.099 | 1.68 | 1.04 | $2 \times 10^{-5}$ |

**TABLE 1**: Results of linear regression analysis and t-test.